\documentclass[aps,prl,twocolumn,showpacs,superscriptaddress,groupedaddress]{revtex4}  % for review and submission
\usepackage{graphicx}  % needed for figures
\usepackage{dcolumn}   % needed for some tables
\usepackage{bm}        % for math
\usepackage{amssymb}   % for math
\usepackage{latexsym,amsmath}

\begin{document}

% The following information is for internal review, please remove them for submission
\widetext
%\leftline{Version xx as of \today}
%\leftline{Primary authors: Joe E. Physics}
%\leftline{To be submitted to (PRL, PRD-RC, PRD, PLB; choose one.)}
%\leftline{Comment to {\tt d0-run2eb-nnn@fnal.gov} by xxx, yyy}
%\centerline{\em D\O\ INTERNAL DOCUMENT -- NOT FOR PUBLIC DISTRIBUTION}

% the following line is for submission, including submission to the arXiv!!
%\hspace{5.2in} \mbox{Fermilab-Pub-04/xxx-E}

%\title{Highly nonlinear wave propagation in diatomic chain with various mass ratios}% Force line breaks with \\
%\title{Experimental and numerical verifications of nonlinear stress wave manipulation via resonating and anti-resonating diatomic granular crystals}% Force line breaks with \\
%\title{Nonlinear stress wave manipulation via resonating and anti-resonating diatomic granular crystals}% Force line breaks with 
%\title{Experimental verification of resonance phenomenon in nonlinear granular crystal leading to strong attenuation of stress waves} \\
\title{Nonlinear Low-to-High Frequency Energy Cascades in Diatomic Granular Crystals}% Force line breaks with\\

\author{E. Kim}
\author{R. Chaunsali}
\affiliation{Aeronautics and Astronautics, University of Washington, Seattle, WA 98195-2400}
\author{H. Xu}
\affiliation{Department of Mathematics and Statistics, University of Massachusetts, Amherst, MA 01003-4515}
\author{J. Jaworski}
\author{J. Yang}
\affiliation{Aeronautics and Astronautics, University of Washington, Seattle, WA 98195-2400}
\author{P. Kevrekidis}
\affiliation{Department of Mathematics and Statistics, University of Massachusetts, Amherst, MA 01003-4515}

\affiliation{Center for Nonlinear Studies and Theoretical Division, Los Alamos
National Laboratory, Los Alamos, NM 87544, USA}

\author{A.F. Vakakis}
\affiliation{Department of Mechanical Science and Engineering, University of Illinois at Urbana Champaign, Urbana, IL 61822}
% \email{jkyang@aa.washington.edu}

% \homepage{http://www.Second.institution.edu/~Charlie.Author.}
%\affiliation{%
%Second institution and/or address%\\This line break forced% with \\
%}%


\begin{abstract}

We study wave propagation in strongly nonlinear 1D diatomic granular crystals under an impact load. Depending on the mass ratio of the `light' to `heavy' beads, 
this system exhibits rich wave dynamics from highly localized traveling waves 
%with various waveforms 
to highly dispersive waves featuring strong attenuation. We experimentally demonstrate the nonlinear resonant and anti-resonant interactions of particles 
and %explore the corresponding dynamics of the solitary waves in this system. Moreover, we observe 
verify that the nonlinear resonance results in strong wave 
attenuation, leading to highly efficient nonlinear energy cascading without 
relying on material damping.  
In this process, mechanical energy is transferred from low to high frequencies, 
while propagating waves emerge in 
both ordered and chaotic waveforms via a distinctive spatial cascading. This  energy transfer mechanism from lower to higher frequencies and wavenumbers 
is of particular significance towards the design of
 novel nonlinear acoustic metamaterials with inherently passive energy redistribution properties.
% Investigating the resonance mechanism more closely using numerical simulations reveals that there exists a unique way in which energy relocates itself in the system.%
%Valid PACS numbers may be entered using the \verb+\pacs{#1}+ command.
\end{abstract}

\pacs{45.70.-n 05.45.-a 46.40.Cd}% PACS, the Physics and Astronomy

\keywords{}%Use showkeys class option if keyword
\maketitle

{\it Introduction.}
Granular crystals, i.e., periodically packed arrays of solid particles interacting elastically via nonlinear contacts, have recently become popular test beds 
for understanding fundamental structures emerging within 
nonlinear wave dynamics, 
such as traveling waves \cite{Nesterenko01, sen08, Jayaprakash01}, 
shock waves \cite{hong05,doney06,dar06}, discrete breathers \cite{Boechler01, Chong02}, and 
nanoptera \cite{eunho01}. In these systems, tunable energy transport is of particular interest. For example, the wave 
propagation speed can vary significantly 
from linear to highly nonlinear regimes by changing the external 
compression applied to the granular system \cite{Nesterenko01}. The energy 
propagation patterns can be tuned from highly localized to modulated or widely dispersed shapes by introducing local resonances in 
constituents~\cite{eunho01}. The controllable 
degree of disorder also changes the nature of energy attenuation, 
%characteristics with propagating waveforms from solitary-like waves which attenuate exponentially to shock-like waves which attenuate following power-law relationships
enabling both exponential and power-law type 
scenarios~\cite{Ponson01, Upadhyaya01}. 
%Similarly, diffusivity of the energy transport varies from sub-diffusion to super-diffusion by altering the degree of nonlinearity %(tuned by pre-compression) 
%and disorderness of the system \cite{Alejandro01}. 
This wealth of energy transport characteristics stems from
the widely tunable interplay of nonlinearity,
discreteness, and heterogeneity in such media.
%stem from the combined effect of nonlinear interactions among particles (e.g., via Hertizan contact) and the coupling of the propagating waves with localized modes in granular crystals. 

%A recent study highlighted that disordered granular media change from a fluid-like state at zero pre-compression to a solid-like state under high pre-compression \cite{Upadhyaya01}. In the fluid-like state, the granular systems have zero stiffness (called \textit{sonic vacuum} due to zero sound speed) and do not support shear force in a multi-dimensional packing under ideal frictionless conditions. In this regime, the granular crystals show various nonlinear wave dynamics depending on the mechanical properties of constituents and their arrangements \cite{Nesterenko01, eunho01, eunho02, doney06, Upadhyaya01}. For example, when mechanical waves propagate through a heterogeneous chain, energy can be split into multiple packets, resulting in efficient wave attenuation \cite{doney06, hong05, fernando}. %This can lead to the engineering application of granular crystals for impact protection purposes 

Among the different heterogeneous variants of granular crystals, 
an ordered diatomic granular crystal -- in the form of alternating heavy and light particles -- is one of the most prototypical 
ones~\cite{Jayaprakash01,mason,Jayaprakash03}.
Remarkably, this simple \textit{dimer} lattice 
shows highly intriguing energy transport phenomena. Previous studies demonstrated that diatomic granular chains support 
various types of highly localized traveling waves, nonlinear beating pulses, 
and highly dispersive waves, depending on the mass ratio of the heavy and light beads~\cite{Jayaprakash01, Jayaprakash03, mason, betti}. 
%{\bf; in these
%works %(see also~\cite{jayaprakash4,jayaprakash5,jayaprakash6}) 
%the notions of resonance and anti-resonance for the dimer
%lattice has been elucidated. 
 It is of particular interest that wave attenuation can be maximized at specific mass ratios, while efficient transmission of energy without attenuation can also occur at another set of mass ratios, resulting in highly nonlinear solitary
traveling waves. This pertains to the so-called nonlinear \textit{resonance} and \textit{anti-resonance} mechanism~\cite{Jayaprakash01, Jayaprakash03}. 
The mass ratios satisfying 
the resonance and anti-resonance conditions appear in an interlaced manner 
as the relevant parameter is varied.
Despite previous theoretical/numerical studies on  
these resonance and anti-resonance mechanisms, corresponding
experimental work has been extremely limited~\cite{Potekin01}.

%The nonlinear resonance and anti-resonance phenomena in granular crystals have been numerically studied in \cite{Jayaprakash01,Jayaprakash03}, but 

Here, we push considerably forth the experimental state of the art
of the system by enabling a full-field 
visualization of lattice's dynamics via laser Doppler vibrometry.
%, rather than at the end of the chain.
These experimental advances, in turn, lead us to consider crucial
theoretical/numerical aspects, such as the nonlinear energy transfer mechanisms
across length and frequency scales. Specifically, we investigate how the resonances 
chiefly disintegrate and
transfer the energy into stable periodic modes, leading to extremely efficient nonlinear energy scattering
and impact mitigation mechanisms. The remainder of the energy
is partitioned to a wide range of frequencies and modes, resulting in an 
apparently chaotic spatial tail dynamics. This set of features is in sharp contradistinction with the anti-resonance scenario where
the energy appears to localize in traveling wave \textit{quanta} (i.e., 
isolated wavepackets). 

Our approach towards exploring the energy transfer
mechanisms in granular crystals, in addition to substantial experimental developments, is motivated theoretically 
by the usefulness of this type of approaches in revealing cascading
dynamics in a large variety of other systems. These include 
classical and quantum fluids~\cite{holmes,Nazarenko01}, plasmas~\cite{Vedenov01}, and solid structures such as vibrating thin 
plates~\cite{During01,Boudaoud01,Mordant01}. %Given that 
%the temporal and spatial redistribution of mechanical energy in 
%granular crystals is largely unknown, 
Previous studies tackled the topic of energy distribution in granular crystals in the context of tapered and
decorated chains~\cite{sen08, doney06}, 
disordered settings~\cite{hong05},
chains bearing impurities~\cite{Job01},
interaction of waves with their own radiation~\cite{r05},
and that of narrow or broadband frequency signals with waves stemming from reflections~\cite{Hutchins01}. Nevertheless, 
the mechanism of energy cascading across time- and length-scales has not been 
suitably explored yet. We believe that this study will
plant the seeds towards the investigations of such energy cascading dynamics in granular crystals as well as nonlinear lattice systems more generally.

\begin{figure}
\includegraphics[scale=0.45]{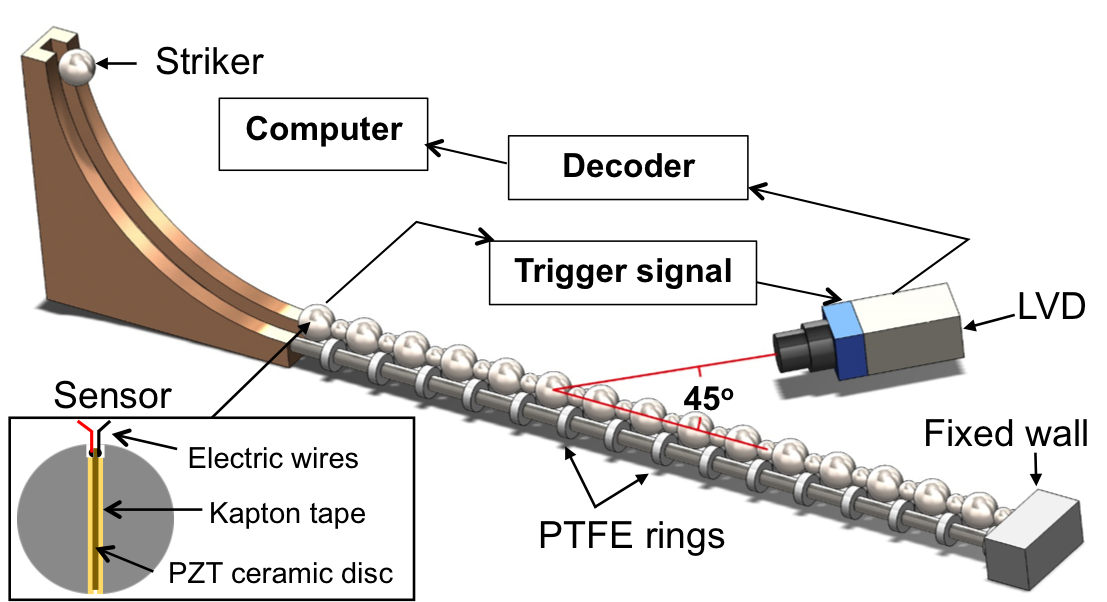}
\setlength{\abovecaptionskip}{3pt}
\caption{Schematic of experimental setup for measuring nonlinear wave propagation in a diatomic granular crystal. Inset shows an instrumented sensor particle.}
\label{fig_dm1}
\end{figure}

\begin{figure*}
\includegraphics[scale=0.58]{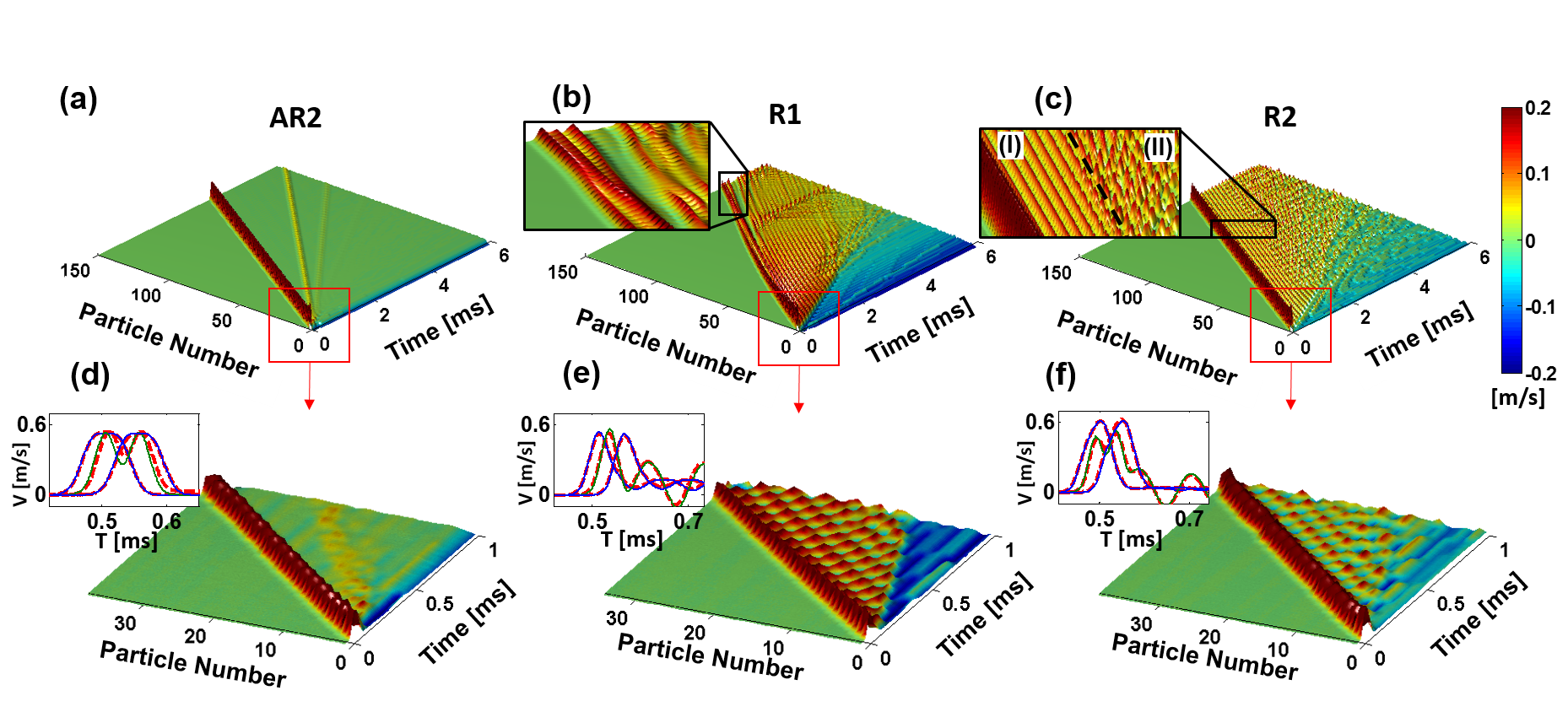}
\setlength{\abovecaptionskip}{-15pt}
\caption{Numerical and experimental surface maps for (a, d) AR2, (b, e) R1, and (c, f) R2. The insets in (b) and (c) show an enlarged view of nonlinear wave beating phenomenon in R1, and regular (I) and irregular (II) forms of wave tail in R2, respectively. % The insets in (d-f) illustrate comparisons of light particle's velocity profiles between numerical (solid blue curve) and experimental (dashed red curve) results.
The insets in (d-f) illustrate comparisons of three successive (heavy-light-heavy) particles'�� velocity profiles between numerical (solid blue(heavy) and green(light)) and experimental (dashed red) results.} 
\label{fig_dm2}
\end{figure*}
\
{\it Experimental and numerical setup.}
Figure 1 shows the test setup consisting of a granular chain, a striker, and a laser Doppler vibrometer (LDV). The chain is composed of alternating `heavy' and `light' spherical particles. All particles  are made of chrome steel ($E = $ 210 GPa, $\nu$ = 0.29, $\rho$ = 7,810 kg/m$^3$). The radius of the heavy particles is $R$ = 9.525 mm. We alter the radius of the light particles to set various mass ratios for the chain (see Supplemental Document [24] for details). We position the smaller particles on polytetrafluoroethylene (PTFE) rings tied to stainless steel rods to align the center of mass of all particles (Fig.~\ref{fig_dm1}). We generate nonlinear stress waves in the granular crystal by impacting one end of the chain with a striker, which is identical to the heavy particle. The striker is released from a ramp and attains an impact velocity of %$v_{stk}$ = 
0.94 $\pm$ 0.017 m/s, hitting the first bead which is
a heavy bead with an embedded piezoelectric ceramic disk at its center (see the inset of Fig.~\ref{fig_dm1}). At the time of impact, this sensor bead generates high voltage, which triggers our data acquisition system. The other end of the chain is fixed with a heavy steel wall.

We position the LDV at a slanted angle (45$^o$ in this study) to measure the velocity of a particle in each impact event (Fig.~\ref{fig_dm1}). We repeatedly measure the particle's velocity by shifting the LDV along the guiding rail and synchronize the measured data with respect to the striker's impact moment to reconstruct the wave propagation profiles. 
We examine the propagation of waves in two anti-resonance cases, labeled AR1 and AR2 
with mass ratios of 1 (the homogeneous case, operating as a
benchmark) and 0.34 respectively, and for two resonance cases, labeled R1 and R2 
with mass ratios of 0.59 and 0.24 
respectively~\cite{Jayaprakash01,Jayaprakash03}.

We complement these experimental results by numerical simulations with a discrete element model. The chain is modeled 
with point masses, which are connected by nonlinear springs representing the Hertzian contact between spheres~\cite{Johnson}. Therefore, the equations of motion of the diatomic granular chain can be written as follows:
\begin{eqnarray*}
\label{eqn1_1} 
M_{2n-1}\ddot{u}_{2n-1}=&\beta[u_{2n-2}-u_{2n-1}]_+^{3/2}-\beta[u_{2n-1}-u_{2n}]_+^{3/2},
\\
\label{eqn1_2} 
M_{2n}\ddot{u}_{2n}=&\beta[u_{2n-1}-u_{2n}]_+^{3/2}-\beta[u_{2n}-u_{2n+1}]_+^{3/2}.
\end{eqnarray*}

\noindent Here $M$ and $u$ represent the mass and displacement of particles (odd and even subscripts correspond to heavy and light beads), $\beta$ is the contact coefficient between two particles~\cite{SI}, and $[s]_+$=max$(s,0)$, implying that the system does not support tensile force. We note that the dynamics of this system is fully re-scalable with energy, so for sufficiently small impulsive excitations the results do not change qualitatively~\cite{Jayaprakash01,Jayaprakash03}.

{\it Results and discussion.}
Figure~\ref{fig_dm2} shows numerical and experimental surface maps of the wave propagation in AR2, R1, and R2. In Fig.~\ref{fig_dm2}(a), 
%multiple solitary waves are formed in the transition region by multiple collisions between light and heavy particles. 
the impact leads to the formation of a number of robust solitary pulses
(three of these pulses are directly visible), which \textit{quantize} the energy repartitioning
it primarily into the first one~\cite{SI, Job02, sen06}. % among them, with considerably smaller parts allocated to the secondary and tertiary solitary wave (which, in turn, 
%per their commensurately smaller amplitude, have also a commensurately smaller speed).
The experimental results in Fig.~\ref{fig_dm2}(d) obtained by the LDV corroborate the numerical simulations, though the tertiary packet is not distinguishable due to the dissipation in experiments. It should also be noted that Figs.~\ref{fig_dm2}(a) and (d) employ different time and length scales because of the shorter granular chain used in experiments.
In this anti-resonance case, the traveling waves propagate with 
almost constant speeds without 
shedding energy behind the leading pulses. The velocities of these wave 
packets follow the relationship, $V \sim  F_{m}^{1/6}$, where $F_{m}$ is 
the peak amplitude of the dynamic force. This is the 
same as the conventional solitary 
wave in a homogeneous chain (i.e., AR1), which shows a single hump with its 
wavelength of $\sim$5 particles' diameters \cite{Nesterenko01}. In AR2, the wavelength is wider than that in AR1 (see Supplemental Material~\cite{SI}), and the heavy and light particles exhibit different velocity profiles \cite{Jayaprakash01}. We experimentally verified the waveform in AR2  as shown in the inset of Fig.~\ref{fig_dm2}(d).

In sharp contrast to anti-resonances, resonances 
cause the propagating wave to shed its energy behind in the form of 
oscillating tails, thereby attenuating the leading pulses 
(see Figs.~\ref{fig_dm2}(b) and (c) for R1 and R2, respectively) \cite{Jayaprakash03}. We 
accurately measured the particle velocity profiles in these cases as 
well (see Figs.~\ref{fig_dm2}(e) and (f)). In R1, the primary 
wave loses a considerable fraction of its energy by 
forming a wave tail oscillating initially in a regular fashion. However, 
after approximately 80 particles, three leading peaks form a wave packet 
(see the inset in Fig.~\ref{fig_dm2}(b)). These pulses exchange energy among them as the wave packet propagates in the medium. This behavior is 
attributed to the so-called nonlinear beating phenomenon \cite{Jayaprakash03}. 
We observe that once the nonlinear beating starts, the energy transfer from the leading wave packet to the wave tail is substantially reduced, as the relevant waveform \textit{detaches} itself from the rest
of the radiative tail. 
In R2, however, the primary wave continuously loses its energy to the wave tail.
In this case, as well as in that of higher order resonances, the oscillations
 of the wave tail retain a very regular structure just behind the primary wave. 
They eventually transform to apparently chaotic forms 
within the far-field [see region (I) and (II) in the inset in 
Fig.~\ref{fig_dm2}(c)].

\begin{figure}
\includegraphics[scale=0.38]{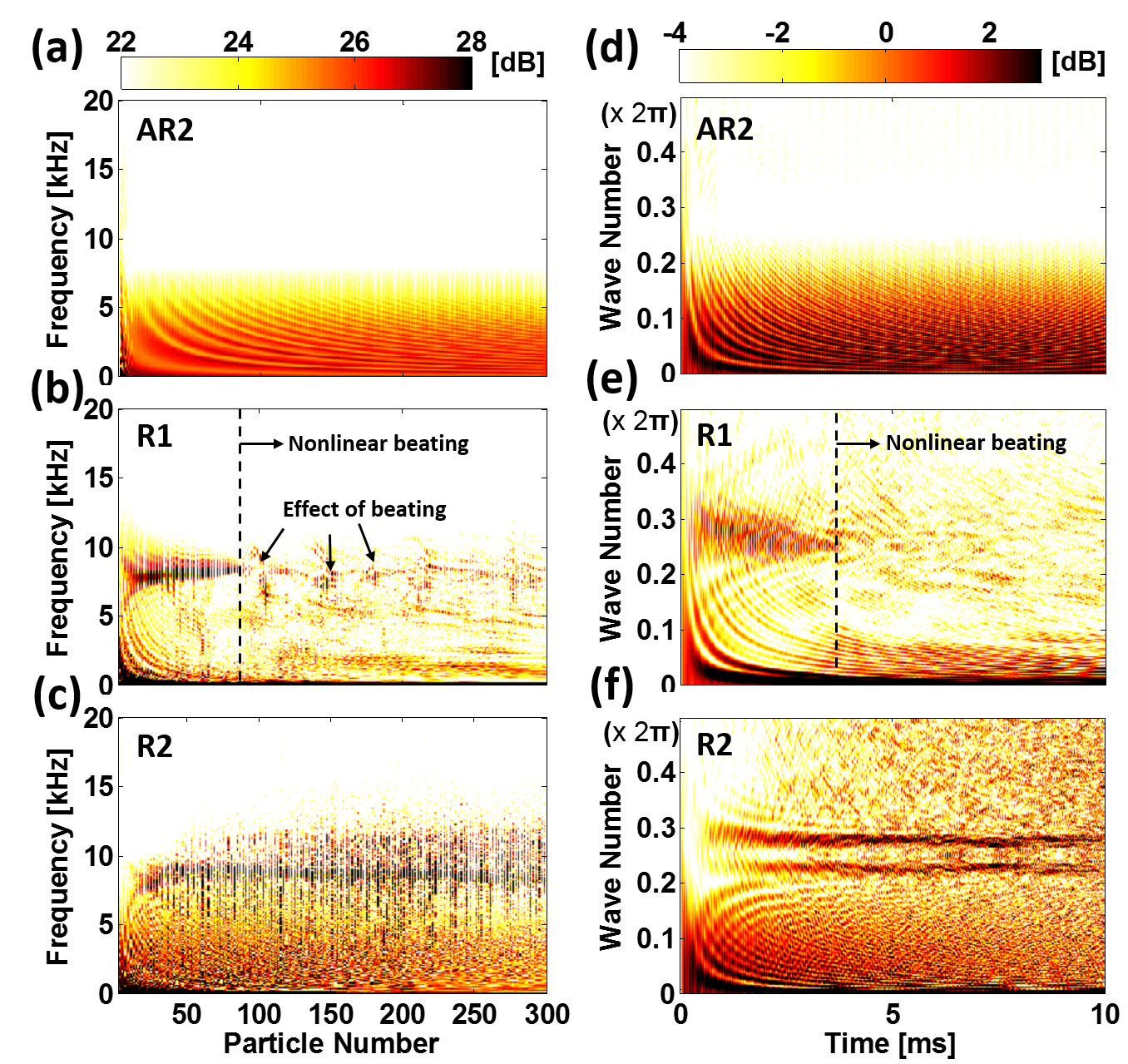}
\setlength{\abovecaptionskip}{-8pt}
\caption{ Frequency and wavenumber spectra of particles' velocity profiles in (a, d) AR2, (b, e) R1, and (c, f) R2, obtained from the analysis of
the computational results and illustrating the presence of cascades
associated with the redistribution of energy among different scales. The wavenumber is calculated based on the wavelength 
normalized by the average diameter of the large and small particles.}
\label{fig_dm4}
\end{figure}

To further investigate the energy dispersion mechanism, 
we consider the relevant phenomenology in the Fourier domain. 
Figure~\ref{fig_dm4}(a) shows the frequency spectrum of the particles' velocity profiles in AR2, where the majority of energy is concentrated in low frequencies. In R1 (Fig.~\ref{fig_dm4}(b)), however, we find that a strong high frequency signal 
($\sim$9 kHz) appears, and the magnitude of the low frequency 
signal decreases during the primary pulse's shedding its energy to 
the wave tail (around up to 80 beads). This predominant mechanism of 
energy transfer ceases to exist as soon as the nonlinear beating starts. This represents the fact 
that the mechanical energy carried initially by the low-frequency primary pulse is redistributed to the higher-frequency contained in the wave tail due to the nonlinear resonance mechanism, but is subsequently trapped in the beating mechanism
without further substantial decay. 
In R2, this energy transfer from low to high frequencies happens persistently (see Fig.~\ref{fig_dm4}(c)) owing to the continuous energy shedding to the wave tail. %as shown in Fig.~\ref{fig_dm2}(c). 
We can discern in the relevant 
frequency patterns the presence of regular oscillations in the wave tail 
[corresponding to part (I) in the inset of Fig.~\ref{fig_dm2}(c)].
Additionally, 
there exists a fraction of the energy deposited to a wide range of frequencies
corresponding to a dispersion of the energy
in other modes, resulting in the apparently chaotic oscillations of section (II) within the inset of 
Fig.~\ref{fig_dm2}(c).
%, the high frequency energy disperses to a wide spectrum of frequency in an irregular form (see the speckles in Fig 3(b)). 

Equally telling as the frequency-domain responses, are the 
wavenumber plots %as a function of time %(for different times, pasted together) 
in Figs.~\ref{fig_dm4}(d-f).
%spatial length scales of the nonlinear waves also change when the energy disperses with the resonance mechanism. 
In contrast to AR2, the resonance cases reveal how the energy of the primary wave at 
%a large length-scale (
small wavenumbers is converted to a wavenumber around $\pi/2$ in the wave 
tail [see Figs.~\ref{fig_dm4}(e) and (f) in comparison to 
Fig.~\ref{fig_dm4}(d)]. 
Remarkably, 
%this smaller length-scale corresponds to the sum of diameters of the 
%heavy and light beads. We note 
this length scale corresponds to the length of two sets of heavy and light beads. We note
that the two strong signals near $\pi/2$ in R2 are because of the Fourier transform of mixed signals from the heavy and light beads in an alternating pattern (see Supplemental Document~\cite{SI}). 
This wavenumber trend suggests the emergence of a definitive 
{\it periodic traveling wave} pattern that
is being excited in association with a periodicity of two-beads.
Such periodic patterns are witnessed experimentally in Figs.~\ref{fig_dm4}(e) and (f), and they have also been studied
analytically in dimer chains~\cite{betti}. Our wavenumber analysis reveals the excitation of these
states in a transient way for R1, until the beating pattern forms. In R2, contrarily, the pattern is persistent 
and clearly discernible in region (I) of Fig.~\ref{fig_dm2}(c),
enabling the continuous transfer of energy from the principal
wave to the associated background.
Once again in R2, when the wave tail transforms to chaotic oscillations, 
the length scale also spreads out widely as denoted by
the speckles in Fig.~\ref{fig_dm4}(f).

\begin{figure}
\includegraphics[scale=0.48]{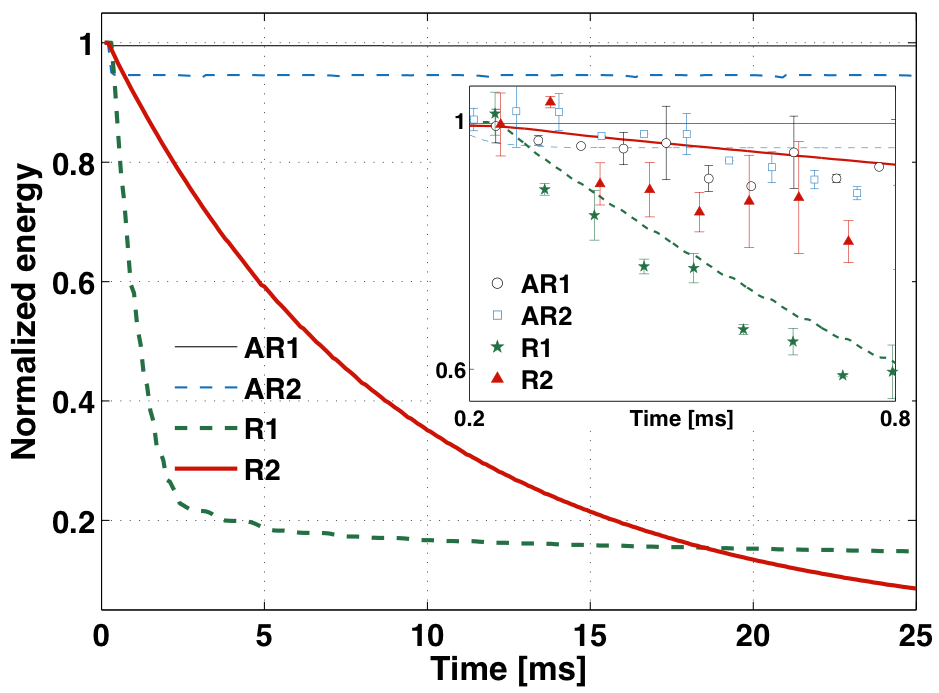}
\setlength{\abovecaptionskip}{-2pt}
\caption{Normalized energy carried by the primary wave packet in anti-resonances and resonances. Inset shows the comparison between the experimental and numerical results. The error bars represent standard deviations based on 5 tests.}
\label{fig_dm3}
\end{figure}

An alternative diagnostic towards the characterization of 
the energy transmission is provided in 
Fig.~\ref{fig_dm3} under 
various anti-resonance and resonance conditions. Here, we compare the energy carried by the leading 
wave packet after normalizing it with respect to the impact energy injected 
by the striker~\cite{SI}. We observe 
highly efficient and constant energy transmission in AR1 and AR2 
due to the formation of solitary pulses. The normalized 
energy is less than 1 in AR2, which is caused by the trailing solitary waves 
generated in the transition region as explained above. In sharp 
contrast to such anti-resonance cases, the primary pulse experiences 
a fast and dramatic attenuation in R1. About 80\% of the energy is transferred 
to the wave tail within just 3 ms, which corresponds to the time the wave takes to propagate through the first 80 particles (see Fig. 2(b)). 
Hence, the emerging periodic traveling wave 
is extremely efficient in rapidly draining the energy of the leading
pulse. However, once the 
nonlinear beating pulse is formed, the transfer of energy to the wave tail is 
essentially suppressed. On the other hand,
in R2, the reduction of normalized energy is slower compared to that in R1. 
Nonetheless, such wave attenuation happens continuously given the persistence
%(as we saw previously) 
in the latter case of the periodic traveling wave.
As a result, the normalized 
energy carried by the primary pulse approaches asymptotically to zero. 
This eventually leads the attenuation of the primary pulse in R2 
to become larger than that in R1. Higher order
resonances share the principal characteristics of R2, as concerns the
above phenomenology. 

The distinctive energy scattering phenomena between R1 and R2 stem from the difference in relative motions of the heavy and light beads. When the particles are squeezed by the leading wave, light particles move faster than heavy particles. The disparity between the heavy and light particles' motions becomes more evident as the mass ratio decreases. %Also, in resonance conditions the energy scattering becomes faster as the light mass increases. 
In R1, the light and heavy beads move at relatively similar frequencies due to the small difference in their masses (mass ratio of 0.59). This results in the fast energy scattering under the facilitated transition of energy between the heavy and light particles. %Interestingly in this case, after rapid scattering of pulses, the resonance mechanism turns into another state showing energy balancing among three pulses called `nonlinear beating' as mentioned above. 
In R2, however, the light particles' motions are much faster than those of heavy particles due to the small mass ratio (only 0.24). This results in a slow -- but more consistent -- reduction of energy, leading to the higher wave attenuation performance than R1. Given the limited length of the granular crystal tested in this study, we obtained experimental data up to $\sim$0.8 ms (inset of Fig. 4). We find that the experimental results corroborate the numerical simulations. 

These observations provide us with a complete picture 
-- in both real and Fourier space, in both time and frequency domains --
of how
wave attenuation mechanisms can be achieved on the basis of 
nonlinear resonance mechanisms. This type of 
process is fundamentally 
different from the conventional energy attenuation mechanisms relying on 
material damping and/or structural deformation effects. Here, the wave attenuation is achieved by maximizing the energy dispersion and redistribution within the chain of granular particles. The existence of damping in this system 
only enhances the wave attenuation. This is confirmed by our 
experimental data, where the energy of the primary pulse attenuates more than that in numerical simulations (Fig.~\ref{fig_dm3}). Moreover, the energy cascading from low to high frequency and large to small length scale 
plays a crucial role in a way partially reminiscent of similar cascading
phenomena within fluidic systems. %when damping exists in the system, because high frequency energy will be eventually dissipated via damping. %This interesting energy cascading observed only in nonlnear system which is not common in the wave propagation in solid structures. wave attenuation based on nonlinear resonance mechanism can offer a new way to design impact protecting layers 

{\it Conclusions.}
In the present work, we experimentally and numerically investigated the nonlinear resonance/anti-resonance phenomena in ordered diatomic granular 
crystals, and their connection to highly effective nonlinear frequency redistribution and cascading of energy across different scales. We verified the existence of efficient energy transfer mechanisms for resonances, and robust
traveling solitary waves for anti-resonances.  
%We presented a complementary (to earlier theoretical/numerical studies) yet fundamental (in its own right) approach focusing on the dynamics of energy transmission/transfer through the chain.
In particular, we demonstrated that the nonlinear resonance mechanism can be highly useful in attenuating impact energy by dispersing it to the wave tail. 
In this process, we found an interesting energy transition mechanism, 
i.e., the 
low-to-high frequency energy transfer and large-to-small scale energy cascading. 
These cascades resemble analogous cascades encountered in turbulent flows, which shows that acoustic metamaterials such as the granular diatomic chain considered herein can be tuned to possess this type of turbulence-like behavior.
A dominant role within this mechanism was revealed to be played by periodic
traveling waves which appear to be excited either 
transiently (R1) or permanently (R2).
This can offer a new way to designing nonlinear acoustic metamaterials based on 
inherently passive energy redistribution principles for impact mitigation purposes.  
 
%The variations of inter-particle 
%stiffness as well as of the particle mass in different can be explored to trigger resonance and anti-resonance mechanisms, which will be reported in the our upcoming publications. 
{\it Acknowledgements.}
We thank Matthew Toles for helping graphical illustrations. JY acknowledges the support of NSF (CMMI-1414748) and ONR (N000141410388). 
%\textbf{Panos and Alex, please add.}
PGK gratefully acknowledged support from the US-AFOSR under grant
FA9550-12-10332. PGK's work at Los Alamos is supported in part by
the U.S. Department of Energy. AFV would like to acknowledge the support of MURI grant US ARO W911NF-09-1-0436.

\end{document}

% The following information is for internal review, please remove them for submission
\widetext
%\leftline{Version xx as of \today}
%\leftline{Primary authors: Joe E. Physics}
%\leftline{To be submitted to (PRL, PRD-RC, PRD, PLB; choose one.)}
%\leftline{Comment to {\tt d0-run2eb-nnn@fnal.gov} by xxx, yyy}
%\centerline{\em D\O\ INTERNAL DOCUMENT -- NOT FOR PUBLIC DISTRIBUTION}

% the following line is for submission, including submission to the arXiv!!
%\hspace{5.2in} \mbox{Fermilab-Pub-04/xxx-E}

\title{SUPPLEMENTAL MATERIAL:\\
Nonlinear Low-to-High Frequency Energy Cascades in Diatomic Granular Crystals}

\pacs{45.70.-n 05.45.-a 46.40.Cd}% PACS, the Physics and Astronomy

\maketitle

\section{I. Hertzian contact and total energy of each particle } 
In the diatomic granular chain, contact interaction between two beads can be expressed by the Hertzian law as follows:

\begin{equation}
F=\beta[u_{n}-u_{n+1}]_+^{3/2}
\label{eqn3}
\end{equation}

\noindent where $\beta$ is a contact coefficient which depends on material properties and geometry of the two objects in contact. In case of a contact between large and small beads made of an identical material, the coefficient is defined as follows:

\begin{equation}
\beta=\frac{2E}{3(1-v^2)}\sqrt{\frac{R_{L}R_{S}}{R_{L}+R_{S}}}
\label{eqn4}
\end{equation}

\noindent where $E$ and $v$ are Young's modulus and Poisson's ratio of the beads respectively, and $R_{L}$ and $R_{S}$ are radius of the large and small beads, respectively~\cite{Johnson}. In the case of the last bead with radius $R$ is in contact with flat surface of the wall made of identical materials with the bead, its contact coefficient becomes:

\begin{equation}
\beta=\frac{2E\sqrt{R}}{3(1-v^2)}
\label{eqn5}
\end{equation}

The equation of motions of the granular chain shown in the manuscript are solved numerically using the Runge-Kutta method with a $10^{-7}$s time step. For this numerical simulation, we use a 300-particle chain --although only a fraction of these, typically $150$ are shown in simulations e.g. of Fig.~2--. The numerical simulation shows accuracy of about 0.004\% error in terms of the total energy conserved during a 20 ms period. It should be noted that we do not consider damping effects in the numerical simulation in order to focus on the wave attenuation caused solely by energy re-distribution in the chain.
If we assume that the potential energy between the two particles in contact is equally distributed to the two particles, the total energy of the $n$-th particle at time $t$ can be defined as follows:

\begin{eqnarray}
E_n(t)&=&\frac{1}{2}M_{n}\dot{u}^2_{n} + \frac{1}{5}\beta [u_{n-1}-u_{n}]_+^{5/2}+\frac{1}{5}\beta [u_{n}-u_{n+1}]_+^{5/2}
\label{eqn6}
\end{eqnarray}

\noindent To calculate the energy of the leading wave in Fig. 4 of the main manuscript, we conduct the summation of the energy for seven particles, which are wide enough to cover the envelop of the leading pulse in AR1, AR2, and R2 cases. In R1, however, we add the energy of nine particles to account for the total energy of the beating waveform. The total energy of the leading waves is normalized with respect to the initial excitation energy, $E_0=\frac{1}{2}M_0v_0^2$, where $v_0$ is the impact velocity of a striker.

\begin{figure}
\includegraphics[scale=0.48]{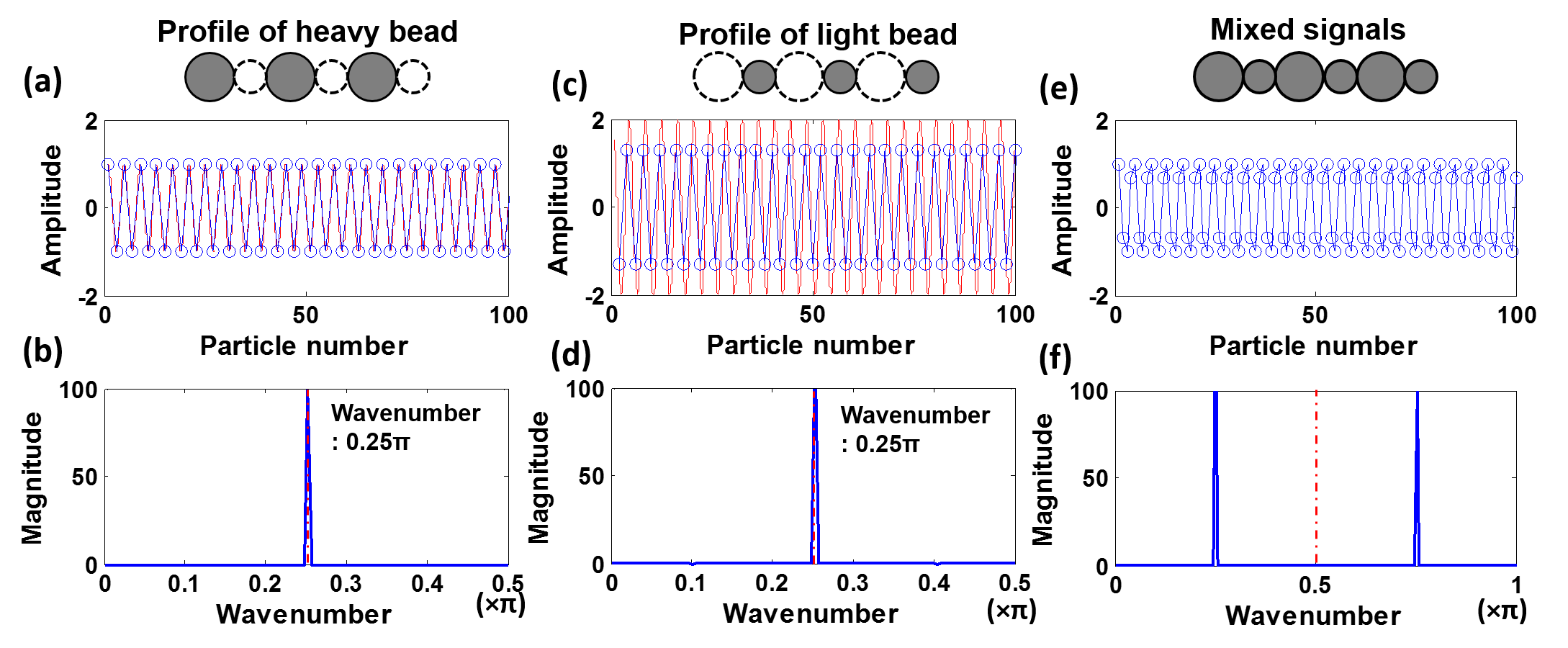}
\caption{Wavenumber analysis in the case that two different profiles with an identical wavenumber of $\pi/2$ are mixed together in an alternating way. Based on the FFT analysis, (b), (d), and (f) show wavenumber of the wave profiles in spatial domains for (a) heavy beads, (c) light beads, and (e) the mixed data of the two, respectively. In (a) and (c), red curves represent continuous signals, while blue circles denote the sampled data at particle positions.}
\label{SM_Fig1}
\end{figure}

\begin{figure}
\includegraphics[scale=0.48]{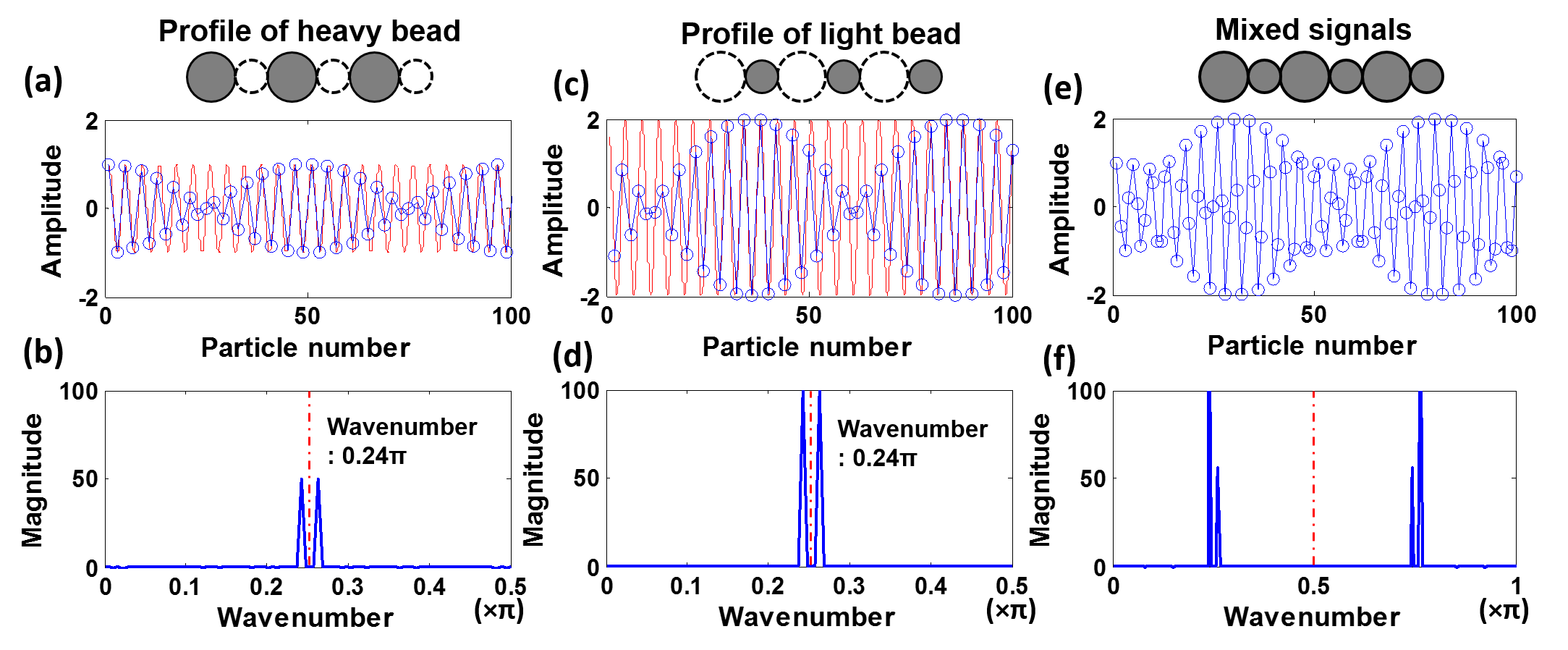}
\caption{Wavenumber analysis in the case that two different profiles with an identical wavenumber of $0.48\pi$ are mixed together in an alternating way. Based on the FFT analysis, (b), (d), and (f) show wavenumber of the wave profiles in spatial domains for (a) heavy beads, (c) light beads, and (e) the mixed data of the two, respectively. In (a) and (c), red curves represent continuous signals, while blue circles denote the sampled data at particle positions.}
\label{ExpSetup}
\end{figure}

\section{II. Parameters of granular crystals in experiment } 
In experiment, we measure wave propagation in two anti-resonance models, AR1 and AR2, and two resonance models, R1 and R2. Their dimensions and mass ratios are summarized in Table I.

\begin{table*}
\label{tab:param}
\caption{Mass ratio and chain length of each experiment models}
\begin{ruledtabular}
\begin{tabular}{cccccc}
Model ID & Large bead&Small bead &Mass ratio& Chain length [mm] \\
             & diameter [mm] & diameter [mm] &  &    (number of beads)  \\
\hline
AR1 & 19.050&19.050&1 & 609.6 (32)\\
R1   & 19.050&15.978&0.59 &595.5 (34)\\
AR2 & 19.050&13.333& 0.34 & 615.3 (38)\\
R2   & 19.050& 11.839& 0.24 & 605.9 (39)\\
\end{tabular}
\end{ruledtabular}
\end{table*}
\

\section{III. Wavenumber of velocity profiles in the first (R1) and second (R2) resonances} 
The high wavenumber components in the nonlinear resonance R1 and R2, shown in Fig. 3(e) and 3(f) respectively, correspond to the \textit{regular} wave tail discussed in Fig. 2 (b) and (c), where `heavy' and `light' beads oscillate in an ordered fashion in the near field of the primary pulse. When we perform Fast-Fourier-Transformation (FFT) of velocity profiles of all beads (i.e., both heavy and light ones) in the spatial domain, we observe one or more wavenumbers symmetric with respect to {\bf$\pi/2$}, which corresponds to the wave length identical to the 4 particles' distance. Here, in the calculation of wavenumber we use the number of particles, instead of their actual dimensions for the sake of simplicity. The symmetry with respect to {\bf$\pi/2$} is due to the FFT of a mixed signal composed of two different velocity profiles. For example, the heavy particles show small amplitude oscillations, while the light particles show larger amplitude ones due to the smaller inertia. If we separate the signals into two profiles, `heavy' particle's velocity profile and `light' particle's velocity profile (see Fig. 5 (a) and (c)), the FFT of these signals show wavenumbers up to {\bf$\pi$}, because the lattice constant is two particles in this case (see Fig. 5 (b) and (d)). These signals are symmetric with respect to the central wavenumber {\bf$\pi/2$} (see Fig. 5 (b) and (d)) due to the nature of the FFT of real valued-signals. 
This is also referred to as a spatial aliasing effect. It is noted that one is mirrored image of the other. Therefore, if the wavenumber of the harmonic signal is {\bf$\pi/2$}, we only can see a single peak  at {\bf$\pi/2$} (Fig. 5(b) and (d)). Otherwise the FFT shows two peaks in symmetry with respect to {\bf$\pi/2$} (Fig. 6 (b) and (d)). 

In the original signals where the two velocity profiles are mixed in an alternating pattern, the lattice constant is a single particle. Therefore, the FFT of this signal shows wavenumbers up to {\bf$2\pi$}, and it is symmetric with respect to {\bf$\pi$}. Interestingly, in this FFT signal of the combined data, the symmetric wavenumbers in the separated velocity profiles (in the range of $0$ to {\bf$\pi$}) appear on the top of each other in the same wavenumber range, from $0$ to {\bf$\pi$}. Therefore, when the oscillations of both `heavy' and `light' particles have a single wavenumber of  {\bf$\pi/2$}, only one wavenumber {\bf($\pi/2$)} appears in the FFT of the combined signal within the range from $0$ to $\pi$. On the other hand, when they have a same wavenumber which is other than {\bf$\pi/2$}, double peaks symmetric with respect to {\bf$\pi/2$} appear. If the two signals have different wavenumbers, the FFT signals show four peaks in symmetry with respect to {\bf$\pi/2$}.

In the first resonance (R1), the high wavenumber component is initially dispersed, and then it is focused around {\bf$\pi/2$} as the wave propagates up to about 80 particles. In the second resonance R2, however, two distinctive lines continuously appear as the wave propagates, which represent that the wavenumbers of the heavy and light particles' oscillations are similar to each other with a minor offset from the {\bf$\pi/2$}. Again, this difference between R1 and R2 stems from the difference of relative particles' motions as we explained in the main manuscript.

\section{IV. Effect of striker mass on the formation of solitary wave trains}

In a homogenous chain (AR1), a solitary wave train appears when the striker mass is larger than the particle mass in the chain \cite{Job02, sen06}. However, in a diatomic chain having the mass ratio of light to heavy particle less than 1 (e.g., AR2), solitary wave train always appears regardless of the striker mass. This is because the striker impact results in multiple collisions among light and heavy particles at the beginning of the chain, which, in turn, generates multiple solitary waves. Figure 7 (a) shows peak forces of the solitary wave train in AR2 at various mass ratios of the striker mass ($M_0$) to the heavy mass of the chain ($M$). Here we only show the peak forces larger than 0.2 N. In this chain, the first particle is a heavy bead, and the mass ratio of the light bead ($m$) to the heavy bead ($M$) is 0.34. We observe that even when the striker mass is less than the light mass ($M_0/M$ = 0.17), a solitary wave train is formed and propagates as shown in Fig. 7(b). The inset shows the solitary wave train formed in the transition region. From Fig. 7(a), we also notice that when $M_0/M$ is larger than a unity, the number of peaks in the solitary wave train substantially increases.

\begin{figure}
\includegraphics[scale=0.75]{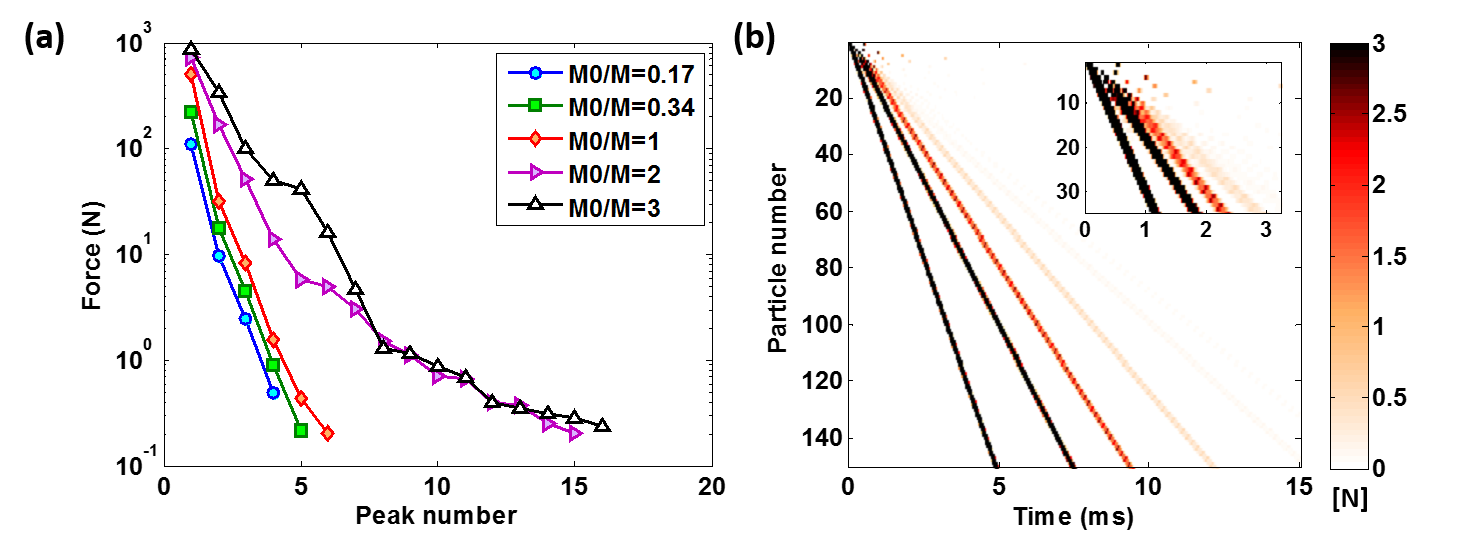}
\caption{Solitary wave train in AR2 at various striker masses (a) magnitude of the peak contact forces for different striker masses ($M_0$) (b) image map of a solitary wave train in AR2 when the striker mass is half of the light mass (i.e., $M_0/M$ = 0.17).}
\label{Wavelength}
\end{figure}

\section{V. Wavelength of the solitary waves in diatomic granular crystals} 
In this section, we compare the wavelength of the solitary waves in diatomic granular chains at various mass ratios. We obtain the wavelength values from three different approaches: (1) the analytical method by Porter \textit{et al.}~\cite{mason} (solid blue curve); (2) the numerical simulations via a discrete element method (green squares); and (3) the experimental results via laser Doppler vibrometry (red circules) as shown in Fig. 8. In the numerical simulation and experiments, we count the number of particles ($n$) having a velocity larger than a threshold value in a fixed time frame, and determine $n+1$ as the wavelength considering the smoothly decreasing wave edges.
%In the numerical simulations and experiments, the wavelengths are calculated by counting the number of particles having a velocity larger than a threshold value in a fixed time frame. 
Here, the threshold is set to be 0.1\% of the maximum particle's velocity. It is to be noted that the wavelength varies depending on the measurement moment due to the discreteness of the system, and it can be overestimated at a small threshold value. Therefore, we process multiple data points at various time frames for statistical analysis. In the analytic expression, the wavelength is about five particles' diameters at mass ratio 1, and it gradually increases and approaches approximately ten particles' diameters as the mass ratio approaches zero. In the experimental and numerical results, the wavelength in AR1 is slightly larger than five particle's diameters. It is generally known that the wavelength of the conventional solitary wave (in AR1) is about 5 particles' diameters. However, it has also been reported that the wavelength in AR1 is close to 7 particles' diameters, if we account for extremely small wave edges \cite{Sen}. This can be captured when we lower the threshold value of detectable wave velocity significantly.

\begin{figure}
\includegraphics[scale=0.6]{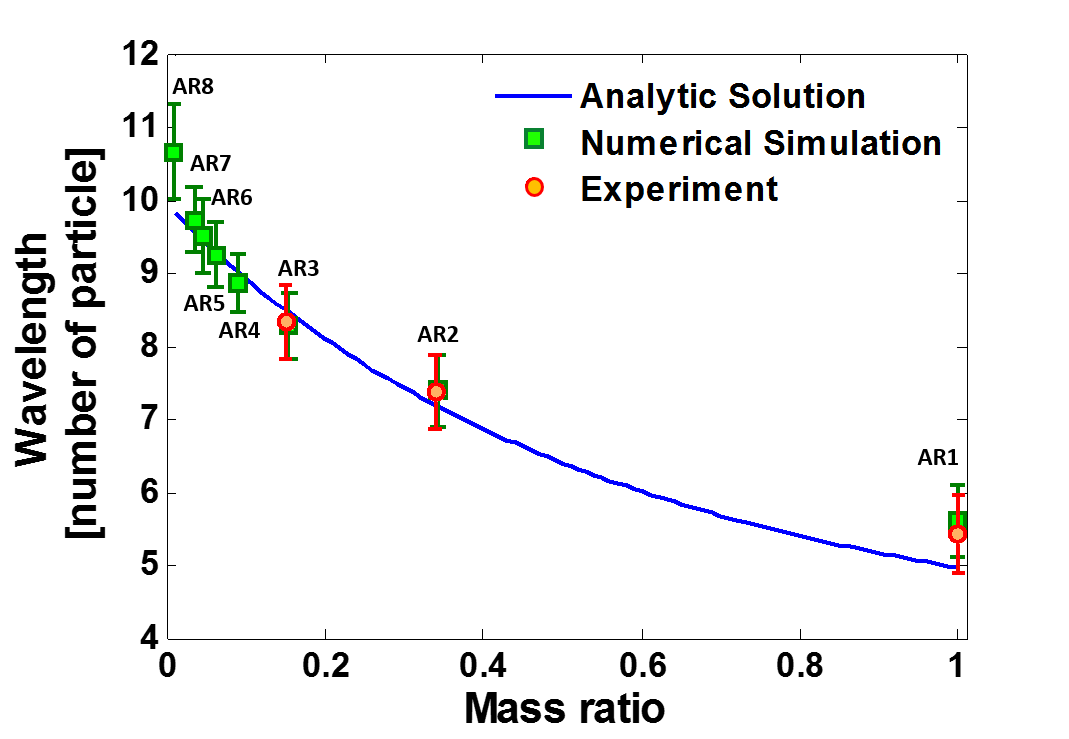}
\caption{Wavelength of the solitary waves in a diatomic chain at various mass ratios from AR1 to AR8. The solid curve represents analytical prediction based on Porter \textit{et al}. \cite{mason}. Green squares and red circles are from numerical simulations and experiments, respectively.}
\label{Wavelength2}
\end{figure}